# Direct-write, focused ion beam-deposited, 7 K superconducting C-Ga-O nanowire

Pashupati Dhakal, G. McMahon, T. Kirkpatrick, J.I. Oh, and M.J. Naughton

Department of Physics, Boston College, Chestnut Hill, Massachusetts 02467

**Abstract**

We have fabricated C-Ga-O nanowires by gallium focused ion beam-induced deposition from the carbon-based precursor phenanthrene. The electrical conductivity of the nanowires is weakly temperature dependent below 300 K, and indicates a transition to a superconducting state below $T_c$ = 7 K. We have measured the temperature dependence of the upper critical field $H_{c2}(T)$, and estimate a zero temperature critical field of 8.8 T. The $T_c$ of this material is approximately 40% higher than that of any other direct write nanowire, such as those based on C-W-Ga, expanding the possibility of fabricating direct-write nanostructures that superconduct above liquid helium temperatures.

Focused ion beam (FIB) techniques have been widely used in the semiconductor industry as well as in nanotechnology research[1,2,3,4,5,6,7,8]. The FIB gas deposition process is similar to chemical vapor deposition, with a source applied in the form of gas precursor via an injection nozzle. A gas injection valve is opened and gaseous molecules from the heated precursor flow and adsorb onto the substrate. The focused ion beam (typically $Ga^+$) is scanned in such as way that it decomposes the adsorbed gas precursor molecules into volatile and non-volatile parts. The former will be deposited on the sample surface, forming the designed structures, while the latter is pumped away. The properties and composition of the deposited materials depend on many parameters, including precursor material, ion beam current, rate of decomposition of the precursor, and distance between the sample and the gas injection nozzle.

The FIB technique has also been used to directly deposit superconducting nanostructures[9,10,11,12,13] (as opposed to using a FIB to shape films down to nanoscale dimensions[14]). Sadki et al.[9] showed that the resistivity of gallium FIB-induced deposition of amorphous tungsten is weakly temperature dependent, similar to a dirty metal but also perhaps close to a metal-nonmetal transition, followed by a superconducting state with $T_c$ between 4 and 5 K. The origin of the superconductivity and its relationship, if any, to a metal-insulator transition in these FIB-deposited films have yet to be elaborated. It is noteworthy that elemental tungsten superconducts, but at millikelvin temperatures in crystals[15] or ~3 K in amorphous form[16], and so the FIB-deposited material appears to be something other than those. Here, we report the occurrence of superconductivity in FIB-induced deposition of a directly-written nanowire from a carbon-based precursor, with a transition temperature of $T_c$=7.0 K, nearly 40% higher than that of the tungsten-based precursor material.

A JEOL Inc. Multibeam JIB-4500 system was used to deposit carbon from a phenanthrene ($C_{14}H_{10}$) precursor onto prepared metallic electrodes on silicon.

Photolithographically-defined microelectrodes (20 nm Ti + 80 nm Au thickness) were evaporated onto a silicon substrate using. The precursor was heated to 85 C and allowed to flow into the chamber using the gas injection nozzle. Upon introduction, the pressure of the sample chamber first increased to $\sim 2\times 10^{-3}$ Pa with the ion beam blanked. Once the pressure of the chamber reached an equilibrium value of $\sim 2\times 10^{-4}$ Pa, the ion beam was set to begin scanning a software-defined area, and deposition commenced. The Ga ion dose used for the deposition was 1 nC/μm$^2$ and the ion beam current was 100 pA.

Figure 1(a) shows an electron microscope image of a FIB-deposited nanowire on the Ti:Au electrodes used for subsequent four probe resistance measurements. Under the conditions employed, the FIB process actually sputter-ablates the surface simultaneous to deposition/writing, such that the final sample deposition occurs in a narrow trench, as shown in the atomic force microscope (AFM) image in Fig. 1(b). As a result, the width of the 35 μm long nanowire can only be estimated at ~100 nm. Figure 2 shows the temperature dependence of the resistance of this nanowire between 2 K and 140 K. Measuring ~4 kΩ at room temperature, it showed weak *T*-dependence until the appearance of a superconducting transition at $T_c$ = 7.0K. As shown, this $T_c$ is determined as the temperature at which the resistance falls to 90% of its extrapolated normal state value, *i.e.* $T_c = T(R/R_n = 0.9)$. These data were taken with a current of 100 nA and in zero magnetic field. Similar measurements at 1 nA and 10 nA gave identical results, verifying the absence of self-heating and suggesting a nonfilamentary nature to the superconductivity. The inset shows the behavior of the normal state, obtained by suppressing superconductivity with magnetic field, exhibiting a broad maximum and minimum near 120 K and 30 K, respectively (magnetoresistance was found to be below 1% at 3 K and 9 T, consistent with a dirty metal). This overall *T*-dependence is similar to that observed in FIB-deposited tungsten, both our own and that from Ref. 9. So, in spite of the sample not exhibiting typical metallic behavior, with increasing normal state resistance at low temperature perhaps indicative of a metal-nonmetal transition, a somewhat sharp ($\Delta T_c$~0.5 K) though incomplete superconducting transition is observed. The later aspect may be associated with the reduced dimensionality of the nanowire, and/or the lack of phase coherence due to nanoscale granularity. Finally, there is evidence from these transport data that a superconducting onset may begin at or even above 11 K.

We have measured the resistive upper critical field $H_{c2}(T)$ of this nanowire via $R(T)$ curves for magnetic fields up to 9 T, spaced by 1 T, shown in Fig. 3. From these data, we extract $H_{c2}$ from $T_c(H)$, using the $R/R_n$=0.9 criterion above. We also note that a superconducting onset occurs at a significantly higher temperature than $T_c$, as shown in the inset. Here, we plot the resistance at each field normalized to its peak value, on an expanded scale that clearly shows the beginning of the resistance decrease at each field. The resulting $H_{c2}(T)$ and $H_{onset}(T)$ data are shown in Fig. 4. As shown, $H_{c2}$ is well fit by a standard pair-breaking formula[17] with a zero temperature critical field of 8.8 T, which corresponds to a superconducting coherence length of 6 nm. The field-dependence of the onset temperatures may indicate a fluctuation regime significantly (>50%) above $T_c$, to 11K or more. Due to the amorphous structure and high normal state resistance, any fluctuations seem more likely of Aslamazov-Larkin[18] origin, as opposed to Maki-Thompson[19] clean limit superconductors. Further studies will be required to clarify this issue.

TEM diffraction on such FIB-deposited samples reveals a lack of Bragg peaks, informing that the material is not crystalline but rather amorphous, similar to previous W-deposition studies. Structural analysis was also conducted on the superconducting nanowire, using energy-dispersive X-ray spectroscopy. The atomic concentrations of the nanowire were found to be 34.2

± 4.3% carbon, 37.5 ± 3.5% oxygen and 26.1 ±2.6% gallium. Notably, these are much higher concentrations of gallium and oxygen than were found in our own 4.5 K superconducting W-containing samples, as well as in those reported in the literature. We found that the concentration of oxygen in the present C-based precursor depositions can be increased by post-annealing in the presence of oxygen in a plasma microwave system. We have also found that the resistance of the nanowires decreases with increasing oxygen concentration in the sample. These anomalously large concentrations of oxygen and gallium may have important roles in the electrical conductivity and especially the superconductivity of FIB-deposited carbon.

Superconductivity is well-established in carbon-based compounds such as organic charge transfer salts[20] and doped fullerides[21], and has been claimed as well in various arrangements of reputedly undoped carbon nanotubes[22,23,24,25]. As stated above, focused ion beam-deposited carbon is not pure carbon and is amorphous in nature. The study by Sadki, *et al.*[9] suggested that the superconductivity observed in FIB-deposited tungsten (via $W(CO)_6$) is also affected by the C and Ga contents, which in their case were in the ratio C:W:Ga ~ 40:40:20%. In more elaborated work by Li, *et al.*[11], the $T_c$ of FIB-deposited tungsten varied between 5 and 6 K, while changing the concentration of C, W, and Ga by varying the FIB current. There, C:W:Ga ratios of ~70:25:10% and ~38:40:22% showed similar $T_c$ ~ 5.0 K, whereas a sample with C:W:Ga ~ 53:33:15% had the highest $T_c$ ~ 6.2 K. The concentrations of carbon and gallium in our sample are comparable to the above reported concentrations in W-samples, but the present sample uniquely contains a significant amount (~38%) of oxygen instead of tungsten. Moreover, no oxygen content was reported for those W-containing materials. It is natural to speculate, therefore, on the roles that gallium and oxygen play in the present superconductivity. We have observed, for example, that samples with lower oxygen concentration do not superconduct, suggesting that the existence of superconductivity in the C-Ga-O system, and the enhancement of its superconducting transition temperature over the C-W-Ga, may be associated with the presence of oxygen, perhaps in combination with gallium. However, wires formed by electron beam-assisted deposition (*i.e.* not gallium ion) with W(CO)$_6$ are reported to have atomic concentrations of 16% tungsten, 63% carbon and 21% oxygen, and are nonsuperconducting[10].

Almost all of the polymorphs of gallium, which number[26] as many as 15, superconduct, with $T_c$ ranging from 1.1 K in the only stable bulk form, α-Ga[27], to 8.4 K in quench-condensed amorphous films[28]. There are also reports[23] of new superconducting phases appearing in nanoconfined Ga, with $T_c$ up to 7.1K. However, no reports have been found that discuss superconducting onsets or fluctuations up to the 11 K temperature shown here. Further studies will be required to determine what role, if any, gallium plays in the occurrence of superconductivity both the previous C-W-Ga and the present C-Ga-O FIB-deposited superconductors.

In summary, we report the presence of superconductivity in a direct-write, FIB-deposited carbon-containing C-Ga-O nanowire. The temperature dependence of the normal state resistivity lies between metallic and insulating, and a superconducting transition occurs at $T_c$ = 7 K, with an onset at 11 K. This superconductivity may be due to the large amount of oxygen and/or the presence of gallium in the sample. Further studies will be required to determine its exact origin. Nonetheless, the direct deposition of superconducting thin films and nanostructures operating above 4.2 K using a FIB may be useful for facile, mask-free fabrication of superconducting coils, nano-SQUIDs and superconducting detectors. Further studies using different precursors and ion beams may result in the *in situ* fabrication of superconducting micro- and nanostructures with yet higher transition temperatures.

This work is supported by the National Science Foundation, Grant No. DMR-0605339.

**Figure Captions**

**Figure 1.** (a) Electron microscope image of FIB-deposited superconducting C-Ga-O nanostructure on lithographically defined Ti:Au electrodes on Si substrate. Bracket above the nanowire indicates the region where atomic force microscopy was performed. Scale bar = 10 μm. (b) AFM topography of the deposited nanostructure revealing the nano-trench due to the FIB process that ablated the surface simultaneous to deposition. Scale bar = 1 μm. The depth of the trench is estimated to be about 200 nm (vertical arrow).

**Figure 2.** Superconducting transition in FIB-deposited C-Ga-O nanowire. The temperature dependent resistance is shown from 150 K to 2 K. $T_c$ was determined at 90% of the normal state resistance. The inset shows an expanded view of the normal state resistance, accessed by suppressing superconductivity with a 9 T magnetic field (dash) and overlapping with zero field data. The temperature $T_{onset}$ where the resistance reaches a local maximum, indicating the onset of the superconducting transition, is also indicated.

**Figure 3.** Temperature dependence of the resistance of a C-Ga-O nanowire in applied magnetic fields between 0 and 9 T, showing the systematic suppression of the resistance decrease consistent with a superconducting state. Inset: $R(T)$ normalized to peak values above $T_c$ at each field, indicating a possible onset of fluctuation-induced superconductivity up to 11 K.

**Figure 4.** Upper critical field vs. temperature, $H_{c2}(T)$, extracted from the data in Fig. 3. The black solid line is a fit to the standard pair-breaking model[17]. Zero temperature critical field, $H_{c2}(0)$ and coherence length, $\xi(0)$ are 8.8 T and 6 nm, respectively. Also shown is a set of data

corresponding to the onset temperatures as defined in Fig. 2, indicating what may be a large fluctuation regime, possibly associated with the reduced dimensionality of the nanowire. Blue line is a guide to the eye.

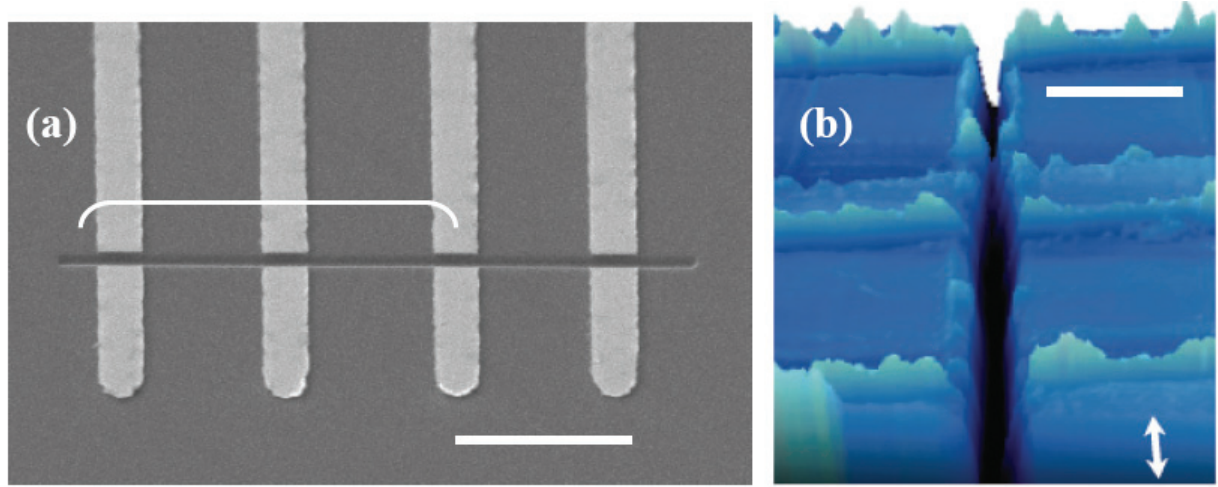

Fig. 1

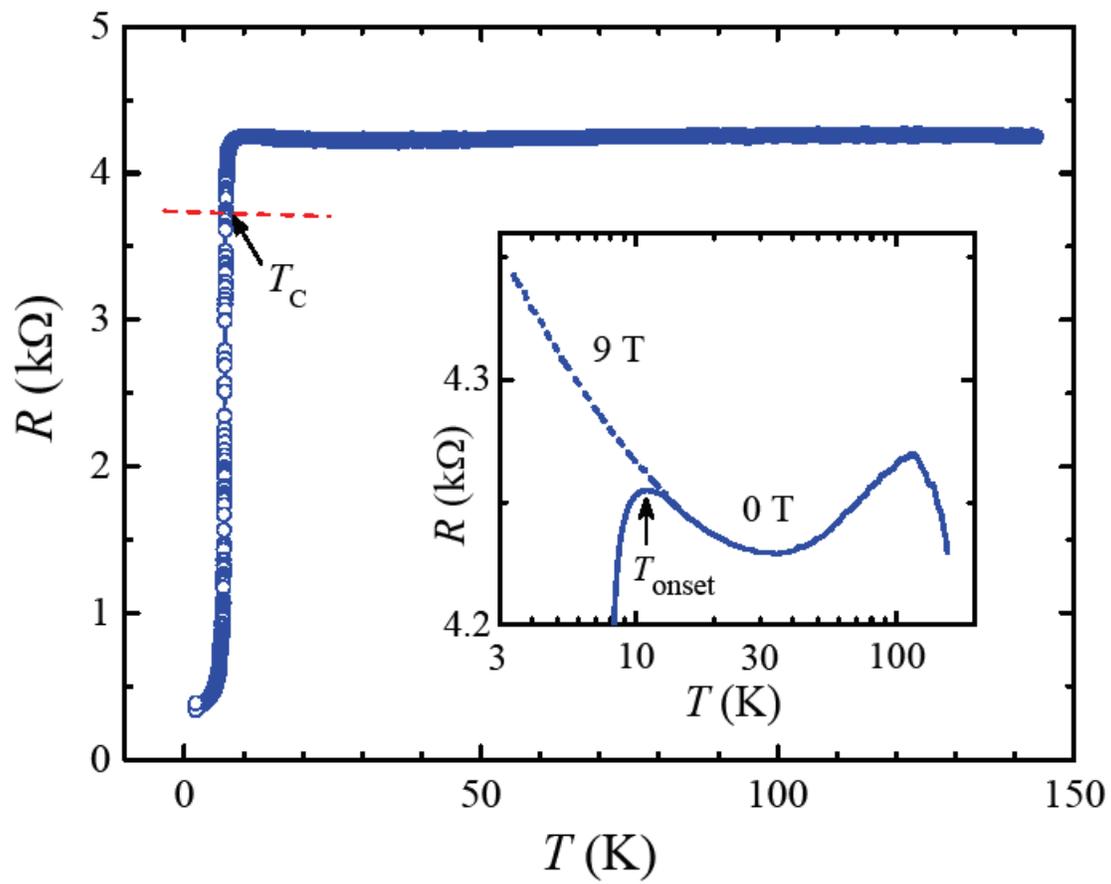

Fig. 2

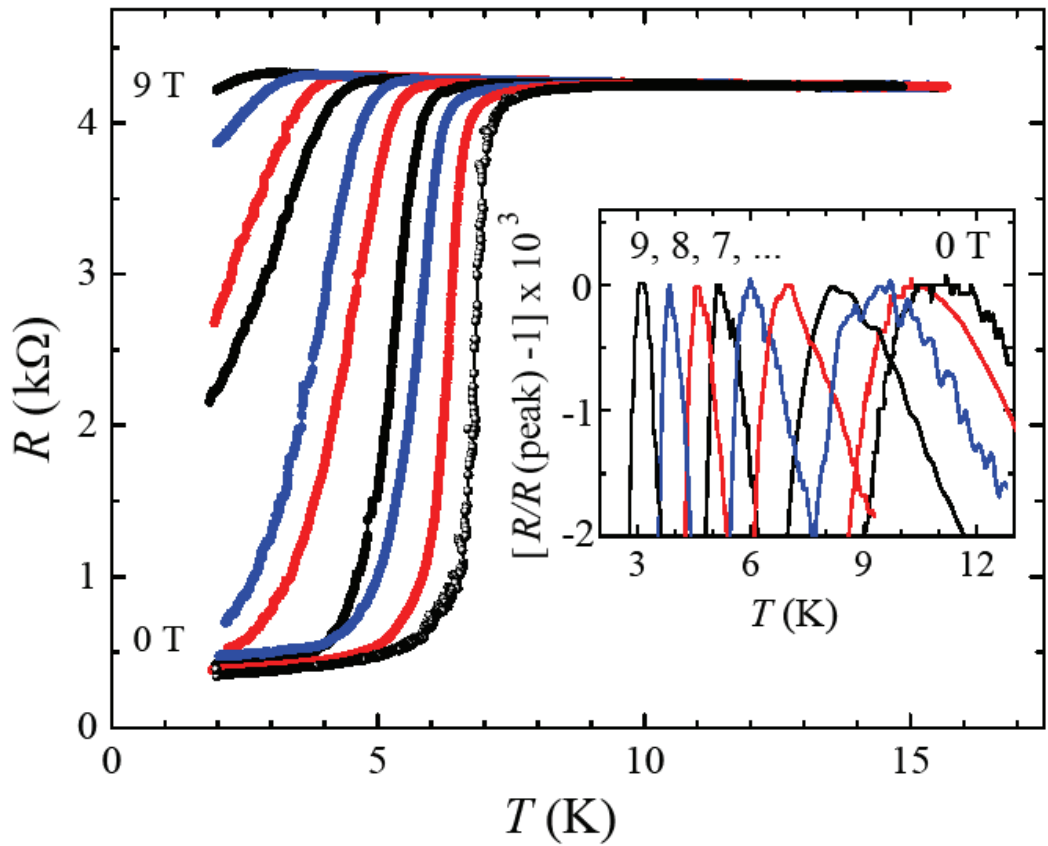

Fig. 3

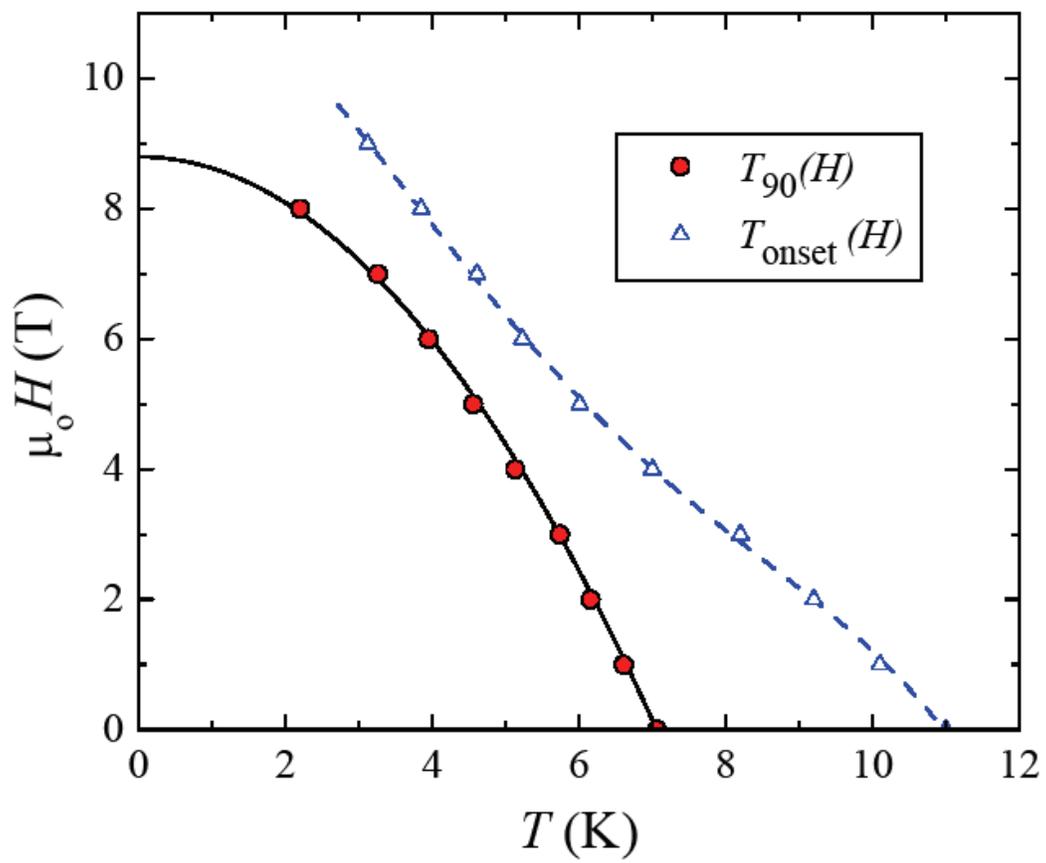

Fig.4